\documentclass[a4paper,12pt]{article}
\usepackage{epsfig}
\usepackage{graphicx}

\newcommand{\Cov}{\mathrm{Cov}}

\begin{document}
\title{Adiabatic theory of  slow extraction and its comparison with non-linear resonance crossing experiment at VEPP-4M} 
\title{Adiabatic theory of  crossing third order  resonance and its comparison with  experiment at VEPP-4M}
\author{S.A. Nikitin, BINP SB RAS, Novosibirsk, Russia}

\maketitle

\begin{abstract}
The universal analytical approach was developed to describe the process of particle
slow extraction from a synchrotron using
adiabatic crossing the betatron resonance of the third order. The obtained formulas make it possible to calculate the time diagrams for the current of the extracted beam  ("spill profile"), taking into account the chromaticity,  particle momentum spread and synchrotron oscillations. On this basis, a method for monochromatization of extracted beam is indicated. The theory is compared  with some results obtained in the experiment on crossing the third-order resonance at the VEPP-4M storage ring.
\end{abstract}

\section*{Introduction}
Since the 1960s, one of the  widely known and used methods for extracting a beam from a synchrotron is the slow crossing of a third-order nonlinear betatron resonance \cite{chao-ref}. In general terms, the theory of the method is based on consideration of the region of stability of particle motion on the phase plane, which decreases as the frequency of betatron oscillations approaches resonance. Different interpretations of this approach are possible (see, for example, \cite{KK,Pullia,NPFP}).
The approach described in this paper with an in-depth analysis of the adiabatic nature of the resonance crossing was developed in the mid-90s, during the author's participation in the research program on obtaining polarized deuterons at the Nuclotron facility (JINR, Dubna) \cite{Dub-ref}.
This approach was previously reported in abbreviated form in \cite{epac2006}. Its distinguishing feature is clarification of the magnitude of the adiabatic interval of motion, which lasts until the rapid increase in the amplitude of particle oscillations.  This is achieved through a rigorous analysis of the phase integral (action integral or adiabatic invariant), which makes it possible to define more precisely in what time a particle with given initial conditions flies out  from a beam.
 We find the adiabatic interval distribution functions for the cases, which differ with respect to accounting the chromaticity,  the particle momentum spread and synchrotron oscillations. Using these distribution functions, formulas are obtained to calculate the time diagrams for the current of the extracted beam  ("spill profile") and the momentum spread of particles in it.  The theory indicates the possibility of obtaining a monochromatic extracted beam due to a small, definitely dosed acceleration of the particles in the process of crossing the resonance. In some numerical examples, we use the Nuclotron parameters from \cite{Dub-ref}.

On the storage ring VEPP-4M, a study was made of particle losses when crossing a third-order nonlinear resonance in vertical motion \cite{JETPH}. 
In this paper, we compare the form of the measured time diagrams of particle losses with those calculated according to the presented theory.

\section{Parameters of motion and non-linearity}

Let list the known relations describing orbital motion of particles in a synchrotron with quadratic nonlinearity of the guide field $\frac{\partial^2 B_y}{\partial x^2}$. The influence of the octupole component of the field, which can stabilize the beam when crossing resonance, is assumed to be negligible.
     Under these conditions, the amplitude of radial betatron oscillations can increase indefinitely near the nonlinear resonance
$$\nu_x=\frac{k}{3}+\delta, \eqno{(1.1)}$$
where $\nu_x$ is the betatron frequency of $x$-oscillations; $\delta<<1$ is the detuning; $k$ is the number of the resonance harmonic of perturbations
 $h(\theta)=
\frac{1}{< B_y>}\frac{\partial^2 B_y}{\partial x^2}$
induced by a system of sextupoles ($< B_y>$ is the  average guide field). 
The function of azimuth $x(\theta)$, which presents radial oscillations of a particle with a zero momentum deviation ($\delta p=0$), is found by the method of averaging and has the Floquet form:
$$x=a_xf_xe^{\imath \nu_x\theta}+c.c..$$
Here, $|f_x|=\sqrt{\beta_x}; $ $c.c.$ is the complex conjugate summand; the amplitude 
$a_x=|a_x|e^{\imath \eta_x}$ is changing 'slowly' by the following law
$$\frac{da_x}{d\theta}=a'_x={\cal P}\bar a_x^2 e^{\imath 3\delta},$$
$${\cal P}=|{\cal P}|e^{\imath  \, arg {\cal P}}=\left<\frac{\imath h}{4}
\bar f_x^3 e^{-\imath k\theta}\right>.$$
The angle brackets mean averaging over the synchrotron storage ring and the bar denotes complex conjugation. In the variables 
$$I_x=\frac{|a_x|^2}{2},$$
$$w=\frac{1}{3}\left(\mbox{arg} {\cal P}-3\eta_x-3\int\limits_0^\theta
\delta d\theta\right)$$
such motion corresponds to the perturbation Hamiltonian $$H=-I_x\delta-\frac{(2I_x)^{3/2}}{3}|{\cal P}|\sin{3w}.
\eqno{(1.2)}$$
Under conditions of slow extraction, Hamiltonian (1.2) depends on the azimuth explicitly since the detuning varies, generally, by the law  
$$\delta=\delta_0-\int\limits_0^{t}\delta' dt.$$

\section{ Perturbation Theory method}

 Applying the expression for the x-motion Hamiltonian near the resonance $3\nu_x = k$ (1.1), we find the phase integral (the action integral)

$$ \Gamma = \oint I dw$$
for the case of the slow changing detuning $\delta  = 3\nu_x - k$:

$$\Bigg |\frac{dH}{d\theta}\Bigg| = |I\delta '| << |H \delta|,$$
where $\delta ' = \frac{d\delta}{d\theta}$. At a distance from the resonance, 
$I = -H / \delta$ and thus this condition takes the following form:
$$|\delta '| << \delta ^2. \eqno {(2.1)}$$
Here $\Gamma$ can be considered an adiabatic invariant ($\Gamma=const$) and one can determine the dependence of the radial oscillation amplitude  
$a_x^2/2 = I$ on $\delta (\theta)$ during slow extraction of particles from a synchrotron. 

At a distance from the resonance, at the initial amplitude of oscillations $I=I_i$ and detuning 
$$|\delta_i|>>\sqrt{2 I_i} |{\cal P}|,$$
a zero approximation is true (${\cal P}\rightarrow 0$)

$$I^{(0)}=<I>=-\frac{H}{\delta} =I_i,$$
$$w'_0 =\frac{d w_0}{d\theta}= -\delta,$$
$$w_0 = -\theta\delta - \frac{1}{3}\mbox{ arg} {\cal P}.$$
$<I> $, a function of $\delta$, denotes the value $<|a_x|^2/2>$ averaged over beatings of the closed phase trajectory $H=const$ in the plane $I,w$. Let call this function the 'action'. Evidently, in a zero approximation $ \Gamma = 2\pi <I>.$
Beatings taken into account, the Hamiltonian $H^{(1)}$  in a first approximation has the following form

$$H^{(1)} = -I^{(1)}\delta(1+\frac{\chi}{\tilde\delta} \sin{3w_0}),\eqno{(2.2)}$$
where
$$\chi = \frac{2}{3}\frac{\sqrt{2 I_i}}{\delta_i} |{\cal P}|\eqno{(2.3)}|$$
is the perturbation parameter (the relative width of the resonance); $\tilde\delta
 = \delta /\delta_i$, $\delta_i$ is the initial value of detuning.
The corresponding amplitude and phase are found from 

$$I^{(1)} = <I> (1-\frac{\chi}{\tilde\delta} \sin 3w_0),$$
$$w_1= w_0 - \frac{\chi}{2\tilde\delta} \cos 3w_0.$$
The phase integral in a first approximation has the following form:
$$ \Gamma = \int\limits_0^{2\pi} I^{(1)} dw_1 =\int\limits_0^{2\pi} I^{(1)}
\frac{dw_1}{dw_0} dw_0 \simeq 2\pi <I>.$$
Moreover, one can directly verify that
$$<I> =-\frac{H_i+\int\limits_0^\theta <I> \delta ' d\theta}
{\delta_i +\int\limits_0^\theta \delta ' d\theta} = const$$
(we have used the relation $\sin 3w_1\simeq\sin 3w_0-\frac{3\chi}{2\tilde
\delta} \cos ^2 3w_0)$. Thus, the first approximation allows one to find the oscillation amplitude 
$$<I> \frac{\chi}{\tilde\delta} =-\frac{H}{\delta}\frac{\chi}
{\tilde\delta} =\frac{2}{3} <I> \frac{\sqrt{2 I_i}}{\delta}|{\cal P}|.$$
The 'action' remains unchanged: $<I>^{(1)} = I_i$. 

In the second approximation,
$$H^{(2)}= -I^{(2)}\delta (1+\frac{2}{3}\frac{\sqrt{2I^{(1)}}}{\delta}|{\cal P}|
 \sin 3w_1),$$
$$w_1 = w_0 - \frac{\chi}{2\tilde\delta} \cos 3w_0,$$
$$I^{(1)}\simeq I_i(1-\frac{\chi}{\tilde\delta} \sin 3w_0). \eqno{(2.4)}$$
Whence
$$I^{(2)} = <I>^{(2)} \left [1-\frac{\chi}{\tilde\delta} \sin 3w_0 +
\frac{\chi^2}{\tilde\delta^2} (\frac{1}{2}+\cos^2 3w_0)\right],$$
$$w_2'=-\delta -\frac{3}{2}\chi\frac{\delta}{\tilde\delta}(1-\frac
{\chi}{2\tilde\delta} \sin 3w_0)(\sin 3w_0 -\frac{3}{2}\frac{\chi}
{\tilde\delta} \cos^2 3w_0),$$
$$ \Gamma = \int\limits_0^{2\pi} I^{(2)} dw_2 = 2\pi <I>^{(2)} (1-\frac{10}{8}
\frac{\chi^2}{\tilde\delta^2})= 2\pi I_i.$$
In the second approximation, the 'action' varies as
$$<I>^{(2)} \simeq\frac{ I_i}{1-\frac{10}{8} \frac{\chi^2}{\tilde\delta^2}}. \eqno{(2.5)}.$$ 
This expression approximately describes how the beat-averaged squared amplitude of x-oscillations grows with decrease of $\delta$ outside the resonance band 
$\delta \sim \sqrt{2I_i} |{\cal P}|$. Below we consider the same effect, using a more accurate  method.  

\section{Phase Integral method}

  In order to analyze in more details the conditions of amplitude growth at the band boundary, we will find an exact expression of the phase integral. Taking the type of symmetry of the trajectories (1.2) at  $H=const$ into account, we can write
$$ \Gamma= 6\int\limits\limits_c^b \frac{I}{dI/dw} dI, \eqno {(3.1)}$$
where $b$ and $c$ are, correspondingly, the maximum and minimum of $I$ at periodic motion. Within the interval $0\leq w\leq 2\pi$ the closed trajectory undergoes 3 full oscillations in accordance with the condition of extremum 
$$\frac{dI}{dw} =\frac{(2I)^{3/2}|{\cal P}| \cos 3w}{-\delta-\sqrt{2I}|{\cal P}|
\sin 3w} = 0, $$
or $\cos 3w=0$. Expressing the dependence of the integrand on $w$ via $I$ and $H$, we obtain
$$ \Gamma = \frac{3}{2\sqrt{2}|{\cal P}|}\int\limits_c^b\frac{(3H+I\delta)}
{\sqrt{I^3-\frac{9}{8|{\cal P}|^2}(H+I\delta)^2}}dI. \eqno{(3.2)}.$$ The radicand in the integrand has a cubic polynomial form and can be presented as 
$$I^3-\frac{9}{8|{\cal P}|^2}(H+I\delta)^2=(a-I)(b-I)(I-c), \eqno{(3.3)}$$
where $a>b>c$ are the real roots of the polynomial. The roots $b$ and $c$ are chosen as the integration limits because of the type of symmetry of closed phase trajectories mentioned above ($c<I<b$), and also because the root $a$ corresponds to the minimum of non-closed trajectories. The latter will be shown below. 

In the mentioned expression, integral (3.2) is a table one and  at $a>b\ge I >c$ equals 
$$ \Gamma=\frac{3}{\sqrt{2}|{\cal P}|}\left\{\frac{(3H+a\delta)}{\sqrt{a-c}}
K(k)-\delta\sqrt{a-c} E(k)\right\}, \eqno{(3.4)}$$
where $K$ and $E$ are complete elliptic integrals of the first and second  kind respectively;
$k=\sqrt{\frac{b-c}{a-c}}$.
Let write (3.4) using the dimensionless variables $\tilde H=H/(I_i\delta _i)$,
$\tilde I = I/I_i$, $\tilde \delta=\delta /\delta_i$, $ \tilde a,\tilde b,\tilde c=a/I_i, b/I_i, c/I_i$, as well as the perturbation parameter 
$$\chi^2 = \frac{8}{9}\frac{I_i |{\cal P}|^2}{\delta_i^2}.$$
We obtain
$$ \Gamma=\frac{2I_i}{\chi}\left\{\frac{3\tilde H+\tilde a\tilde\delta)}
{\sqrt{\tilde a - \tilde c}} K(k) -\tilde \delta \sqrt{\tilde a-
\tilde c} E(k)
\right \}, \eqno{(3.5)}$$
besides $ \Gamma=2\pi I_i$. We will also use (3.3) in a form of equation of extremums:

$$\tilde I^3-\frac{\tilde\delta^2}{\chi^2
}\tilde I^2 - 2\frac{
\tilde H \tilde\delta}
{\chi^2}\tilde I-\frac{\tilde H^2}{\chi^2} = 0. \eqno{(3.6)}$$
The latter expression can be replaced by equivalent combined equations that are derived from the Vieta theorem:
$$\tilde a + \tilde b +\tilde c =\frac{\tilde \delta ^2}{\chi ^2},\ \ \tilde a\tilde b\tilde c =\frac{\tilde H^2}{\chi ^2},\ \  \tilde a\tilde b +\tilde b\tilde c +\tilde a\tilde c=-
\frac{2\tilde H\tilde \delta}{\chi^2}.
\eqno{(3.7)}$$
We will add forth equation (3.5) to these combined equations 
$$-\tilde H =\frac{1}{3}\left[\frac{\pi\chi\sqrt{\tilde a-\tilde c}}
{K(k)}+
\tilde a\tilde\delta-\tilde\delta (\tilde a-\tilde c)
\frac{E(k)}{K(k)}\right].
\eqno{(3.8)}$$
The algebraic system (3.7-3.8) is solved numerically for the decreasing series of detuning values $\tilde \delta$.  For example, by the iteration method with the following initial approximations for the unknowns:
$$\tilde \delta =1;\ \ \ \tilde H=-1;\ \ \ \tilde a=1/\chi^2 - 2;\ \ \  \tilde b=1 + \chi;\ \ \ \tilde c=1 - \chi.$$
Normalized evolution of resonance detuning looks like
$\tilde \delta=1-2\tilde\theta,$
where $\tilde\theta=\theta/(\omega_0 T_{ex})$, $T_{ex}$ is the time interval when detuning changes by the value $2\delta_i$ with the rate $\delta
'=const$. 

Fig.1 shows an example of behavior of the solutions 
$\tilde b(\tilde \theta)$ and
$\tilde c(\tilde \theta)$ at $\chi=0.04$. For the sake of comparison, the values $I_{max}/I_i=1+\chi/\tilde\delta$
and $I_{min}/I_i=1-\chi/\tilde \delta$ derived by the perturbation theory (see Eq. (2.2) at $\sin {3w}_0=\pm1$) method have also been plotted. 
The third root $\tilde a$ (far from the resonance $\tilde a\sim\chi^{-2}>>1$)decreases with detuning, approaching to the growing root
$\tilde b$ (Fig.2). When they finally approach, the discriminant of equation (3.6) changes sign from positive to negative, which means that there is only one real root (and two more complex conjugate roots). 
On the phase plane, the trajectory, which before this critical moment looked like a figure close to a triangle, opens at the trianagle vertices.
\begin{figure}
\centering
\includegraphics*[width=150mm]{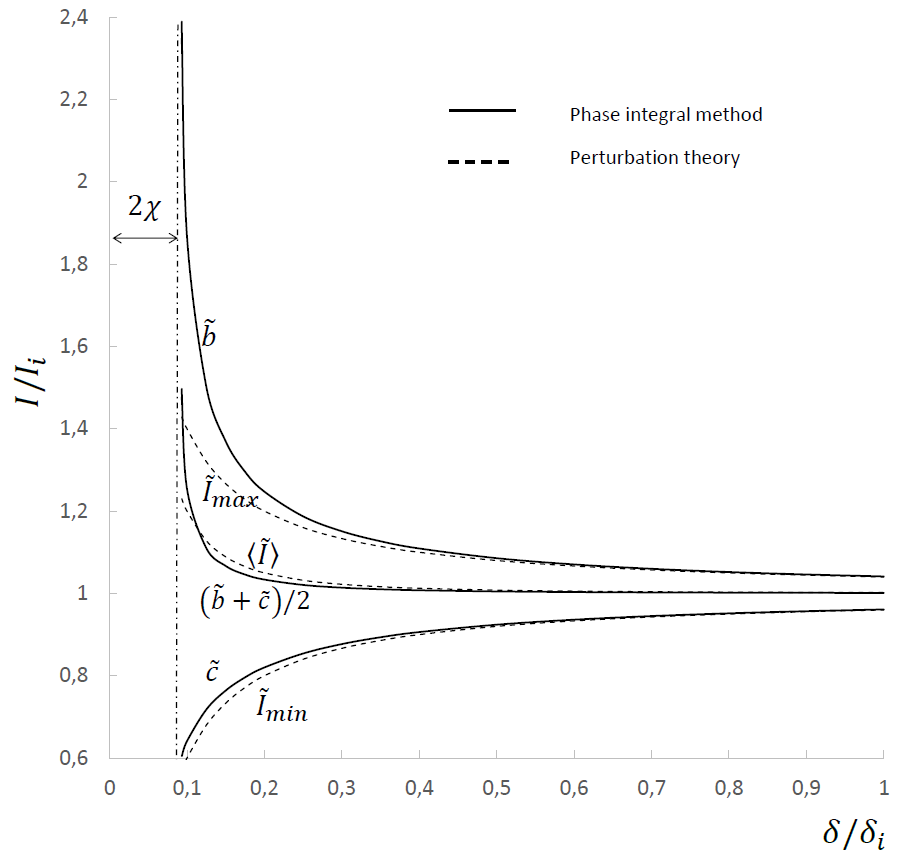}
\caption{{Variation of the maximal, minimal and average values  of the action for closed phase trajectories under adiabatic decreasing the resonance detuning at $\chi=0.04$. The dotted line designates computation in terms of the perturbation theory: $\tilde I_{max}=1+\chi/\tilde \delta$; $\tilde I_{min}=1-\chi/\tilde \delta$; $<\tilde I>=1+10\chi^2/(8{\tilde \delta}^2)$.  Solid lines indicate the use of the exact expression for the phase integral.}}
\label{f1}
\end{figure}
\begin{figure}
\centering
\includegraphics*[width=110mm]{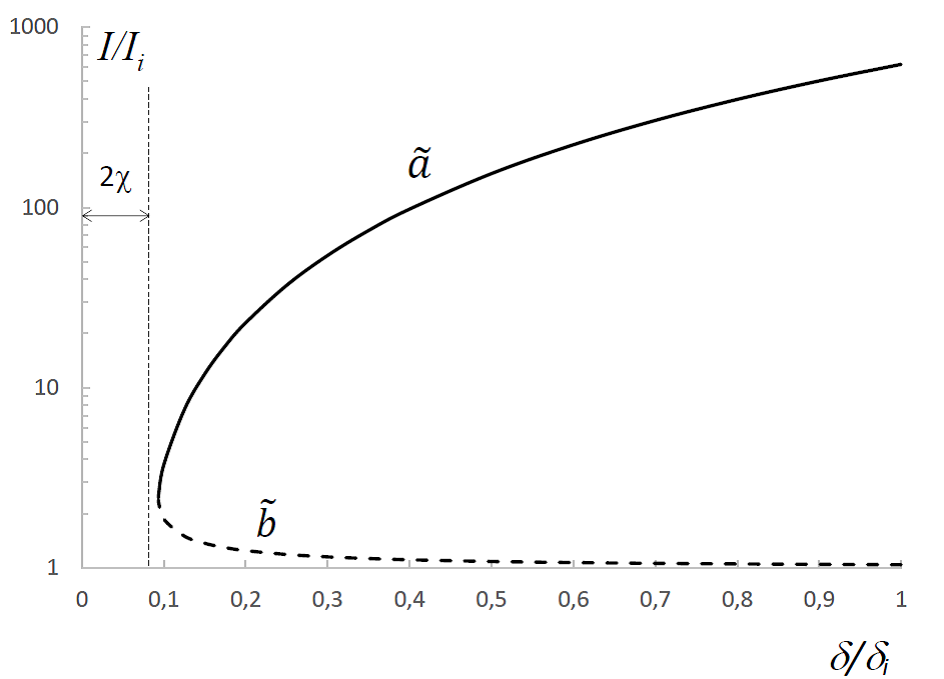}
\caption{ {Convergence of the roots $\tilde a$ and $\tilde b$ with decreasing detuning ($\chi=0.04$).}}
\label{f2}
\end{figure}
\begin{figure}
\centering
\includegraphics*[width=110mm]{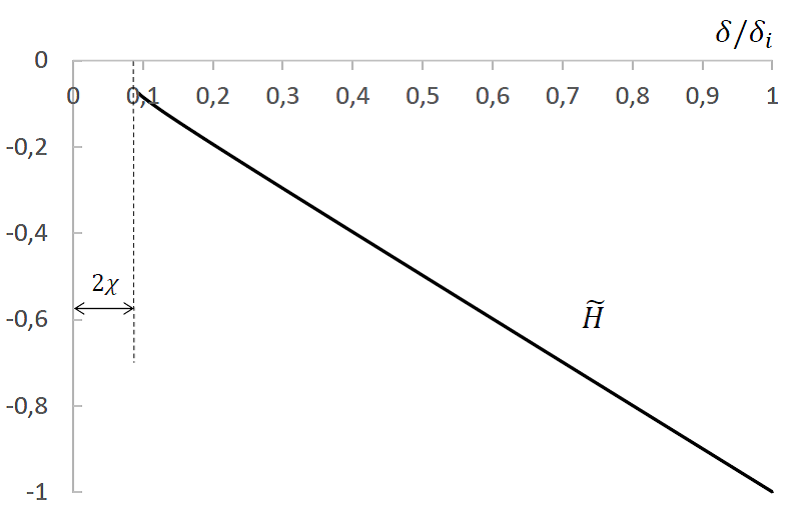}
\caption{{Variation of the Hamiltonian in units of its initial value with adiabatic decreasing rresonance detuning ($\chi=0.04$).}}
\label{f3}
\end{figure}

It follows from the determination of $dI/dw$ (see above) that, with that, it is near the extremum 
$I_e=I_{max}=b$ ($\sin{3w}=-1$) where a singularity arises because of the denominator going to zero: $$-\delta-\sqrt{2 I}|{\cal P}|\sin{3w} \rightarrow 0.$$
Actually, that means that the amplitude $I$ of this interval of phase trajectory increases rapidly while the phase $w$ changes relatively slowly. Otherwise, the condition for such regime of motion to arise can be expressed as
$$1-\frac{3\chi}{2\tilde\delta}\sqrt{1+\frac{\chi}{\tilde\delta}}=0
\eqno{(3.9)}$$
or
$$\chi/\tilde\delta=-1+\sqrt{\frac{7}{3}}\approx0.5.$$
For the plot in Fig.1, in particular, that takes place near the resonance band, whose halfwidth is $\Delta\tilde \delta\approx 2\chi=0.08$. The value
$\chi=0.04$ corresponds to the initial amplitudes   $I_i=\Sigma_x/8\pi$  where $\Sigma_x=2\pi \epsilon_x$  is the circulating beam emittance multiplied by $2\pi$
(in the nuclotron, $\Sigma_x=13\pi\cdot 10^{-4}$ cm$\cdot$rad at 0.2 GeV/nucleon).

Figure 3 shows how the Hamiltonian function calculated in this way in units of its unperturbed value ($I_i\delta_i$) changes in the process of adiabatically slow decrease in the resonant detuning.
It would be interesting to note linearity of this dependence: $\tilde H\propto \tilde\delta $.
It follows from the given results that particles can be extracted slowly before violation of adiabaticity condition (2.1), which as applied to the beam as a whole takes the following form (here,$<..>$ mean averaging over an ensemble of particles): 
$$\delta'<<4\left<\chi^2 \delta_i^2\right>\simeq \frac{1}{9}|
{\cal P}|^2\Sigma_x \eqno{(3.10)}.$$
This conclusion can be easily verified by the given numerical example if the adiabaticity condition is written in normalized variables as 
$$|\tilde I|<<|\tilde H \Delta \psi_b|,$$
where $|\Delta\psi_b|=|\delta_i \tilde \theta|>>1$ is the betatron phase incursion in the time $T_{ex}$. 

Such is an evolution of the picture of motion on the "action-phase" plane. Far from resonance, it is represented by a circle, which, as the detuning decreases, turns into a three-beam star which is close  in shape to a triangle with a closed trajectory.  When crossing the border of the resonance band, the trajectory opens at the vertices of the "triangle".
In the well-known "triangular separatrix" approach, the beam extraction begins when the phase area occupied by it becomes less than the phase area occupied by the beam at a certain critical value of the resonant detuning. For example, in \cite{NPFP} the critical detuning is given by the equation
$$\delta_\Delta\le \frac{1}{4\sqrt{3\sqrt{3}}}|S_k|\frac{|a_x|}{\beta_x},$$
where the  "sextupole harmonic" has a form 
$$ S_k=\frac{1}{2 B\rho}\int_0^L \frac{\partial^2 B_y}{\partial x^2}\beta_x^{3/2}\exp{[- i  k \phi (s)]} ds. $$
Here, $B\rho$ is the magnetic rigity; $L$ is a machine  circumference;  $\phi=\int_0^s ds'/(\nu_x \beta_x)$.  In the smooth approximation $|S_k|/|{\cal P}
|\approx 4\pi$ and  $|a_x|^2\approx{\cal E}_x/4$. In this case, it is easy to compare the value of the critical resonant detuning from \cite{NPFP} with that given in (2.3):
$$\frac{\delta_\Delta}{2\chi\delta_i}=\frac{3}{8\sqrt{3\pi\sqrt{3}}}\frac{|S_k|}{|{\cal P}|}\approx 1.17.$$
Thus, with respect to the specified parameter, the phase integral method is quite consistent with the theory of the "triangular separatrix". 

\section{Adiabatic Interval method}
Adiabatic motion starts  at $\theta=0$, when the excitation is 'switched on' and the resonance detuning begins lessening. This mode ends approximately at $\theta=\Theta$, when the amplitude begins a rapid growth. From (3.9) we find that the adiabatic motion interval  $\Theta$ for an arbitrary particle in a beam meets the equation  
$$2\chi=\tilde\delta=\frac{1}{\delta_i}\left [\delta_i-\int\limits_0^{\Theta}
\left (\delta'-\xi_x\frac{d}{d\theta}\frac{\Delta p_0}{p_0}\right ) d\theta+\xi_x \frac{\delta p}{p_0}\right ],
 \eqno{(4.1)}$$
Here  $\chi=2\rho|{\cal P}|/3\delta_i$, $\rho=|a_x|$; $\delta_i=\delta_0+
\xi_x(u_0+u)|_{\theta=0}$ , the initial detuning, and  $\delta_0$ is its average value in a circulating beam; $\xi_x=\partial{\nu_x}/\partial{u}$
is the chromaticity coefficient; $\delta'=\Delta\nu_x/\omega_0T_{ex}$  is the rate of detuning change in the range $\Delta \nu_x=2\delta_0$.
The term  $u_0=\int\frac{d}{d\theta}\frac{\Delta p_0}{p_0} d\theta $ is a small increment of the average beam momentum relative to the equilibrium one due to a special beam acceleration, which, we assume, occurs at a constant speed $(\Delta p_0/p_0)/\omega_0 T_{ex}$ during the time $T_{ex}$ of the resonant detuning change. In the case of heavy particles, for this, in principle, one can try to switch on an additional accelerating RF resonator during the retuning of the betatron frequency, while the field in the bending magnets does not change. In the electron synchrotron, for this purpose, it is possible to apply a gradual change  in the frequency of the master oscillator of the RF system. The term $\xi_x u=\xi_x \frac{\delta p}{p_0}$ describes the evolution of the detuning of an arbitrary particle, determined by the deviation of its momentum without taking into account the contribution of the total additional acceleration. It is owing to either synchrotron oscillations or, as in the case of  a debunched beam of heavy particles, a "frozen" momentum spread.

\subsection{Zero momentum spread approximation}
In (4.1), the parameters $\delta_i$ and $\chi$ depend on the magnitude of the action $I_i$. Therefore, the particles in the beam will have different values of the  adiabaticity interval $\Theta$ in which they reach the edge of the resonance band.
First of all, we find the distribution function $f_{\Theta}$ of the value $\Theta$ in the beam as well as its first two moments -  the average  $<\Theta>$  and the variance
$<\Theta^2>$ -  in the monochromatic beam approximation 
($\sigma_p=\delta p/p_0=0$). Using (4.1) let express $\Theta$
in relative units for the case of 
$p_0$ linearity ($\Delta p_0'=\Delta p_0/\omega_0T_{ex}$)
$$\tilde \Theta=\frac{\Theta}{\omega_0T_{ex}}=\frac{1}{2q}\left(1-
\frac{4}{3}\rho\frac{|{\cal P}|}{\delta_0}\right), \eqno{(4.2)}$$
$$q=1-\frac{\xi}{2\delta_0}\frac{\Delta p_0}{p_0},\eqno{(4.3)}$$
where $\Delta p_0$ is the magnitude of $p_0$ increment in the time $T_{ex}$.
From (4.2) it follows, in particular, that at $\Delta p_0 \rightarrow 2p
_0\delta_0/\xi$ and $\xi>0$ the real particle extraction is stretched in time compared to the case of $\Delta p_0=0 $.
When determining $f_{\Theta}$ ($\int\limits f_{\Theta}d\Theta=1$),
it would be convenient to express the value $\rho=|a_x|$  in the Twiss parameters \cite{ MSands}
$$\rho=\frac{1}{2}\left
[\frac{x^2}{\beta_x}+\beta_x(x'+\frac{\alpha_xx}{\beta_x})^2
\right]^{1/2},\ \ \alpha_x=-\frac{1}{2}\frac{d\beta_x}{ds}, \ \ \  x'=\frac{dx}{ds};$$
$s=R\theta$, $R$ is the average radius of a synchrotron. Let introduce the characteristic 
$w= 4\rho^2\beta_x,$
proportional to the radial oscillation energy, whose average value is linked with the variances of a trajectory deviation ($\sigma_x^2=<x^2>$)
and a tilt ($\sigma_{x'}^2=<x'^2>$):
$$<w>=4<\rho^2>\beta_x=\sigma_x^2(1-\alpha_x^2)+\beta_x^2\sigma
_{x'}^2. $$

In the smooth approximation ($\beta_x=R/\nu_x, \alpha_x=0,
\beta_x^2\sigma_{x'}^2=\sigma_x^2$)  the probability density function of a random variable $w$ for the gaussian beam is \cite{ MSands}:
$$f_w=\frac{\exp{(-w/<w>)}}{<w>},\eqno{(4.4)}$$
$<w>=2\sigma_x^2$. The probability density function of $\rho$ can be derived from the last equation and has the  Rayleigh's function form:
$$f_\rho=\frac{\rho}{d}\exp{\left(-\frac{\rho^2}{2d}\right)}, \ \ \ d=\frac{\sigma_x^2}{4\beta_x}.\eqno{(4.5)}$$
For completeness we present also the expressions for the mean and variance of $\rho$
$$<\rho>=\frac{1}{4}\sqrt{\frac{2\pi}{\beta_x}}\sigma_x, \ \ \ D\rho=<(\rho-<\rho>)^2>=\left(2-\frac{\pi}{2}\right )d.\eqno{(4.6)}$$

Using the function $f_\rho$ and the definition (4.2)  we obtain the ptrobability density function of the interval $\tilde\Theta$ and then its  mean value and variance:
$$f_{\Theta}=\frac{9\pi\delta_0^2q}{|{\cal P}|^2\Sigma_x}|1-
2\tilde\Theta q|
\exp{\left\{-\frac{9\pi\delta_0^2}{4|{\cal P}|^2\Sigma_x}(1-2\tilde
\Theta q)^2
\right\}},\eqno{(4.7)}$$
$$<\tilde\Theta>=\frac{1}{2q}(1-\frac{1}{3}\sqrt{\frac{2\pi}{\beta_x}}
\frac{|{\cal P}|}{\delta_0}\sigma_x)=\frac{1}{2q}(1-\frac{\sqrt{\Sigma_x}}{3}
\frac{|{\cal P}|}{\delta_0}),$$
$$<(\tilde\Theta-<\tilde \Theta>)^2)>=<\delta\tilde\Theta^2>=\frac{(4-\pi)}{36\pi}\frac{|{\cal P}|^2}{\delta_0^2
q^2}\Sigma_x, \eqno{(4.8)}.$$
Here $\Sigma_x=2\pi\sigma_x^2/\beta_x\approx2\pi\sigma_x^2\nu_x/R$  is characteristic area on the phase plane of radial betatron oscillations (differs from the emittance by the factor $2\pi$) .

Current of particles extracted is determined by the function 
$f_{\Theta}=f_{\Theta}[\tilde\Theta(t)]$
and depends on the time ($t>0$) as
$${\cal J}(t)=eNf_{\Theta}\left |\frac{d\tilde\Theta}{dt}\right |=
\frac{eN}{T_{ex}}f_{\Theta}, \eqno{(4.9)}$$
$e$ is the elementary charge, $N$ is the number of deuterons accelerated. In particular, in the 'smooth' approximation ($\delta p=0$)
$${\cal J}(t)= \frac{eN}{T_{ex}}\frac{9\pi\delta_0^2}{|{\cal P}|^2\Sigma_x}
\left |1-\frac{2tq}
{T_{ex}}\right |\exp{\left\{-\frac{9\pi\delta_0^2}{4|{\cal P}|^2\Sigma_x}
(1-\frac{2tq}{T_{ex}})^2\right\}}.
\eqno{(4.10)}$$
The average time of extraction and its variance equal to $<t_*>=<\tilde\Theta>T_{ex}$ and $\sqrt{<\delta t_*^2>}=
\sqrt{<\delta\tilde\Theta^2>}T_{ex}$.
The extracted beam gets the spread in momentum $$<\delta p_*^2>^{1/2}=\Delta p_0<\delta\tilde\Theta^2>^{1/2}, \eqno{(4.11)}$$
though the circulating beam was monochromatic in the approximation considered. 
In reality such a case could approximately correspond to the condition $\sigma_p<<\Delta p_0$.

Figure 4 presents an example of computation from (4.10) of the time dependence of relative-unit current of extracted particles ($\tilde J=JT_{ex}/eN;
\tilde\Theta=t/T_{ex}$). The spill time structure is plotted
for the following Nuclotron parameters: $E=0.2$  GeV/nucleon;
$\Sigma_x=13\pi\cdot 10^{-4}$ rad$\cdot$cm; $\delta_0=0.035;|{\cal P}|
=0.07$ cm$^{-1/2}$. It is given that
$\Delta p_0=0\hspace{2mm}(q=1)$.

\begin{figure}
\centering
\includegraphics*[width=90mm]{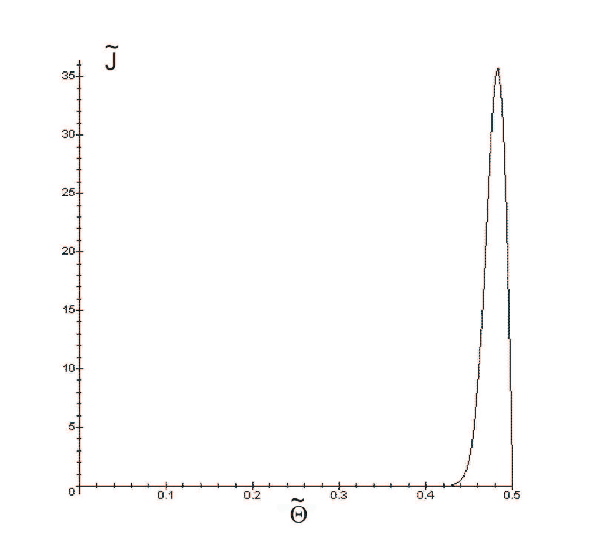}
\caption{ {Extracted particle current $\tilde J={\cal J}T_{ex}/eN$ (in units of average current $eN/T_{ex}$) as a function of time in units of $T_{ex} $ for zero momentum spread in a circulating beam (or zero chromaticity).}}
\label{f4}
\end{figure}

\subsection{Zero chromaticity}

 Practically, the above-considered approximation $\sigma_p=0$ can correspond to the case of zero chromaticity of $x$ -oscillations at a special choice of the system to correct quadratic non-linearity of the guide field ($\xi_x=0$). Then $\Theta$ does not depend on $\delta p$ and variance of momenta in the extracted beam is given by the following expression: 

$$<\delta p_*^2>=\Delta p_0^2<\delta\tilde\Theta^2>+<\delta p^2>
,$$
where $<\delta\tilde\Theta^2>$ is computed with the help of (4.6) and
(4.8) ($q=1$);
$<\delta p^2>^{1/2}=\sigma_p$ is the momentum spread in the circulating beam.

\indent
\subsection{Consideration of momentum spread}
Below, we consider taking into account the momentum spread and synchrotron oscillations for the case of heavy particles, when the effect of radiative diffusion can be neglected.
If the beam is extracted with the RF field shut down (no synchrotron oscillations), the condition for a particle with the momentum deviation $p_0u$ to 'outfly' can be written in the following form:
$$2\chi\approx1-\frac{\delta'}{\delta_i}\Theta, \eqno(4.12)$$
where $\delta_i=\delta_0+\xi_xu$, or ($q\equiv 1$)
$$\tilde\Theta=\tilde\Theta_1+\tilde\Theta_2,\eqno(4.13)$$
$$\tilde\Theta_1=\frac{1}{2q}\left(1-\frac{4\rho}{3}\frac{|{\cal P}|}
{\delta_0}\right),$$
$$\tilde\Theta_2=\frac{\xi_x}{2q\delta_0}\frac{\delta p}{p_0}.$$
The function of distribution $f_1(\tilde\Theta_1)$ coincides with (4.7). The spread in momentum is gaussian, with the variance $\sigma_p^2$: 
$$f_2(\tilde\Theta_2)=\frac{|\Lambda|}{\sqrt{2\pi}\sigma_p}\exp{\left(\frac
{-\Lambda^2\tilde\Theta_2^2}{2\sigma_p^2}\right)}.\eqno(4.14)$$
Spread of the interval value $\tilde\Theta$ can be found via composition of the functions $f_1$ and $f_2$:
$$f(\tilde\Theta)=\int\limits_{\tilde\Theta-1/2q}^
{\tilde\Theta}f_1(\tilde\Theta_1-\tilde\Theta_2)
\cdot f_2(\tilde\Theta_2)d\tilde\Theta_2.\eqno(4.15)$$
Omitting bulky calculation, we will write the equation for current of extracted particles:
$${\cal J}(\tilde\Theta)=\frac{e N}{2T_{ex}}\sqrt{\frac{2}{\pi}}
\frac{\mu |\Lambda|}{\sigma_p(\mu+\tau)}\left\{\sqrt{\frac{\pi}{\mu+\tau}}\tau\ae
\cdot e^{-\frac{\mu \tau \ae^2}{\mu+\tau}}\left[\mbox{erfc}\left(-\frac{\tau\ae}
{\sqrt{\mu+\tau}}
\right)\right.\right.+$$
$$\mbox{}+\left.\left.\mbox{erfc}\left(\frac{\mu+\tau+\tau\ae}{\sqrt{\mu+\tau}}
\right)-2\right]+
e^{-\tau\ae^2}-e^{-\tau(\ae+1)^2-\mu}\right\},\eqno(4.16)$$

Here and next we use the following designations:
$$\Lambda=\frac{2q\delta_0}{\xi_x},\  \  \tau=\frac{\Lambda^2p_0^2}{8q^2\sigma_p^2},\  \  \mu=\frac{9\pi\delta_0^2}{4|{\cal P}^2\Sigma_x},\  \  \ae(\tilde \Theta)=2q\tilde\Theta-1.$$
In this case, spread of momenta in the extracted beam is defined completely by that in the circulating one. 

Figure 5 presents the time diagram  of the current extracted from the Nucletron for conditions in the example in section 4.1. A non-zero spread of momenta $\delta p/p=10^{-3}$ is taken into account at the chromaticity $\xi_x=10$. Other parameters are similar to those from  the example in Figure 4.

\begin{figure}
\centering
\includegraphics*[width=90mm]{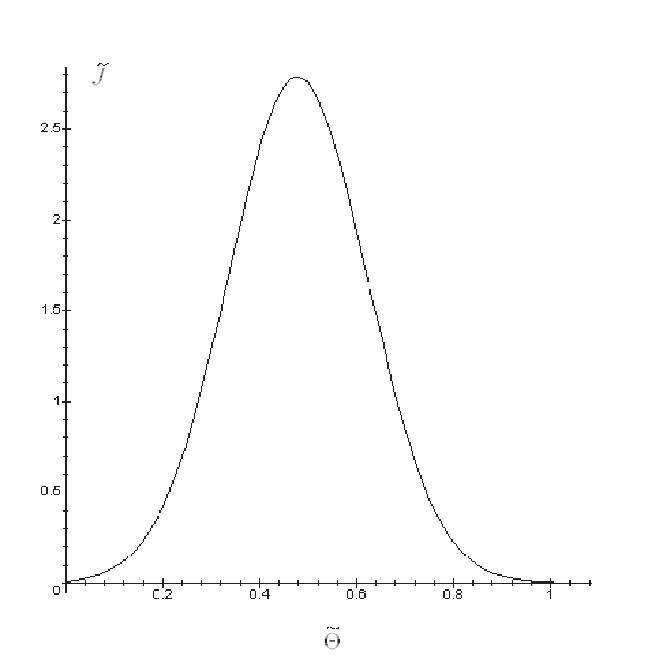}
\caption{{Timing  ($\tilde \Theta=t/T_{ex}$) of the extracted particle current at the momentum spread of $10^{-3}$  in circulating beam  (RF is  shut down).}} 
\label{f5}
\end{figure}

\indent
\subsection{Non-coherent synchrotron oscillations}

Generally ($\xi_x\neq0$), the betatron frequency  
$\nu_x$ is modulated by the law 
$$\nu_x=<\nu_x>+\xi_x \frac{\delta p}{p}\sin{(\nu_s\theta+\phi)},$$
which leads to the splitting of the resonance (1.1) into a series of resonances of the form 
$$3<\nu_x>+m\nu_s=k_m+\delta_m.$$
Here $<\nu_x>$ is the average over the period of synchrotron oscillations $2\pi/\nu_s$; $m$ is an order of the modulation resonance; $k_m$ is a harmonic number;
 $\delta_m$ is a detuning. 
The power of side-band resonances ($m\neq0$)
$${\cal P}_m=(-i)^m J_m\left(\frac{3\xi_x}{\nu_s}\frac{\delta p}{p}\right)
\langle\frac{ih}{4}f_x^{*3}e^{-ik_m\theta}\rangle,$$
$J_m(y)$ is the $m$-order Bessel function.
At a small synchrotron frequency
$$\nu_s<<6\langle\
\chi\delta_i\rangle\simeq|{\cal P}|\Sigma_x^{1/2}, \eqno{(4.17)}$$
the nearest modulation resonances are fully inside the band of the principal resonance (the modulation index $\sim3\xi_x \sigma_p /\nu_s$ can be of the order of 1 and larger). 
At an energy of 0.2 GeV/nucleon  in the Nuclotron this requires 
$$\nu_s<<4.5\cdot10^{-3},\hspace{3mm}f_s=\nu_s\omega_0<<19 \mbox{kHz}$$
($\omega_0=4.2\cdot10^6$ rad/s).  For comparison, in the Nuclotron working energy range, $f_s<10^2 \mbox{Hz}$  ($\nu_s<2\cdot10^{-5}$).
So, for slow extraction with RF voltage switched on, consideration of momentum spread is limited to assessment of the corresponding initial spread of detuning from resonance $\xi_x\delta p/p_0$. Since $\nu_s$ is small, the modified condition of adiabaticity is met\footnote{In electron synchrotrons, the frequency $\nu_s$ is much higher as compared to proton or ion machines, due to which the conditions (4.17) and (4.18) may not be satisfied.}:
$$|\delta'|+|\xi_x\nu_s\sigma_p|<<4\langle\chi\delta_i\rangle^2. \eqno{(4.18)}$$
At $\sigma_p\neq0$ particles 'escape'  when ($\Delta p_0'=
$const, $\delta'=$const)
$$2\chi\approx1-\frac{\delta'}{\delta_i}\Theta+\frac{\xi_x}{\delta_i}\frac{d}{d\theta}\frac{\Delta p_0}
{p_0}\Theta,\eqno{(4.19)}$$
where $\delta_i=\delta _0-\xi_x|\delta p/p_0|$, or 
$$\tilde\Theta=\frac{1}{2q}\left(1-\frac{4\rho}{3}\frac{|{\cal P}|}
{\delta_0}\right)-\frac{\xi_x}{2q\delta_0}\biggl|\frac{\delta p}{p_0}
\biggr|.\eqno{(4.20)}$$
In this case, the modulus of momentum deviation (i.g. an amplitudue value of synchrotron oscillations) implies a positive initial detuning ($\delta _0>0$) and, simultaneously, a positive chromaticity ($\xi_x>0$). Besides, the frequency of synchrotron oscillations should be significantly larger than the inverse time of passing the resonance band with a width of the order of $\chi$. These features constitute an approximation in which we propose to take into account synchrotron oscillations. With the chosen sign of the initial  detuning and chromaticity, it is the amplitude of these oscillations, together with the changing values of the detuning and $\Delta p_0$, that determines the moment when the critical resonance band is reached.

We write (4.4.3) as a sum $\tilde\Theta=\tilde\Theta_1+\tilde\Theta_2$. The distribution function $f_1(\tilde\Theta_1)$ coincides with the similar one found for the case of the spread of momenta of the debunched beam. The distribution function  $f_2(\tilde\Theta_2)$ taking into account an amplitude $|dp/p_0|$ differs from (4.3.3) only by a  factor of 2:
$$f_2(\tilde\Theta_2)=\frac{2|\Lambda|}{\sqrt{2\pi}\sigma_p}\exp{\left(\frac
{-\Lambda^2\tilde\Theta_2^2}{2\sigma_p^2}\right)}.\eqno(4.21)$$
Applying the composition of distribution functions
and the definition (4.9), we now find the expression for the spill structure in consideration of synchrotron oscillations: 
$${\cal J}(\tilde\Theta)=\frac{e N}{T_{ex}}\sqrt{\frac{2}{\pi}}
\frac{\mu |\Lambda|}{\sigma_p(\mu+\tau)}\left\{\sqrt{\frac{\pi}{\mu+\tau}}\tau\ae
\cdot e^{-\frac{\mu \tau \ae^2}{\mu+\tau}}\left[\mbox{erfc}\left(-\frac{\tau\ae}
{\sqrt{\mu+\tau}}
\right)\right.\right.+$$
$$\mbox{}-\left.\left.\mbox{erfc}\left(\frac{\mu\ae}{\sqrt{\mu+\tau}}
\right)\right]+
e^{-\tau\ae^2}-e^{-\mu\ae^2}\right\}. \eqno{(4.22)}$$

Unlike the case of the RF field shut down ($\nu_s=0$, Fig.5), the obtained spill profile 
is asymmetric at similar other parameters (see the left curve in Fig.6). 
\begin{figure}
\centering
\includegraphics*[width=90mm]{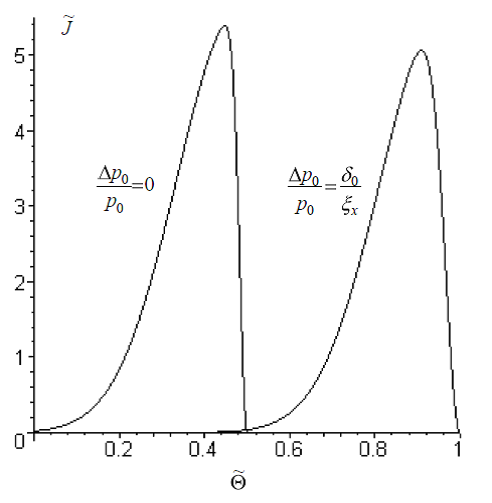}
\caption{ {The spill profile with taking into account the synchrotron oscillations in two cases: without ($\Delta p_0=0$) and with a definitely dozed acceleration  ($\Delta p_0=p_0\delta_0/\xi_x$).  The initial momentum spread is $10^{-3}$; the chromaticity $\xi_x=10$.}}
\label{f6}
\end{figure}
Under these conditions, a dispersion of momenta in the extracted beam is found from the expression
$$<\delta p_*^2>={\cal D}p_*={\cal D}|\delta p|+
(\Delta p_0)^2{\cal D}\Theta_*+2\Delta p_0 \Cov\{\Theta_*,|\delta p|\}.$$
The symbols  ${\cal D}...$  and $\Cov\{...,...\}$ mean a variance and a covariance, respectively. The value $\Theta_*$ is defined by the condition of 'outfly' (4.19). 
As a result, we obtain the relation for the momentum spread of extracted particles:
$$<\delta p_*^2>\approx \frac{\pi-2}{\pi}\sigma_p^2\left(1-\frac{\Delta p_0}
{p_0\Lambda}\right)^2, \eqno{(4.23)}$$
where $\Lambda=2q\delta_0/\xi_x$. As seen, the  presence of slight acceleration ($0<\Delta p_0/p_0 <\Lambda$) at non-zero chromaticity 
($\xi_x>0$) is advantageous for minimization of the momentum spread of the extracted beam\footnote{In this case, apparently,  the extraction should be carried out in the area with zero dispersion. }.  The amplitude of the synchrotron oscillations and the acquired additional momentum averaged over the beam are related by the condition that the edge of the resonance band is reached. With positive values of the initial detuning, chromaticity and additional acceleration, the greater the amplitude, the lower the current increase in the average momentum, and the earlier this event occurs. 
But  the lower the amplitude, the greater the current increase in the average momentum, the critical band is reached later.  In both cases, at the moment of crossing the border of the band, the particle energy is the sum of two mentioned components. 
In principle, a monochromatic extraction could be possible at exact equality  $\Delta p_0 =p_0\Lambda$.  The latter can be alternatively written using the definition for $q$ (4.12) like
$$\frac{\Delta p_0}{p_0}=\frac{\delta_0}{\xi_x}. \eqno{(4.24)}$$
In Fig.7, a particle with an amplitude of synchrotron oscillations $u_1$ reaches the boundary of the critical resonance band, ahead of a particle with a smaller amplitude $u_2$. If (4.4.8 )  is satisfied, the second particle at the band boundary will have a momentum increment equal to the amplitude difference $$\Delta u=u_1- u_2=\frac{\Delta p_0}{p_0}\frac{\Delta\delta}{\delta_0}.$$  As a result, the momentum of each particle is $p_0+\Delta p_0$.  It should be kept in mind that we neglect the contribution of the additional acceleration during the motion of particles inside the resonant band, since it occurs in a relatively short time.
  \begin{figure}
\centering
\includegraphics*[width=90mm]{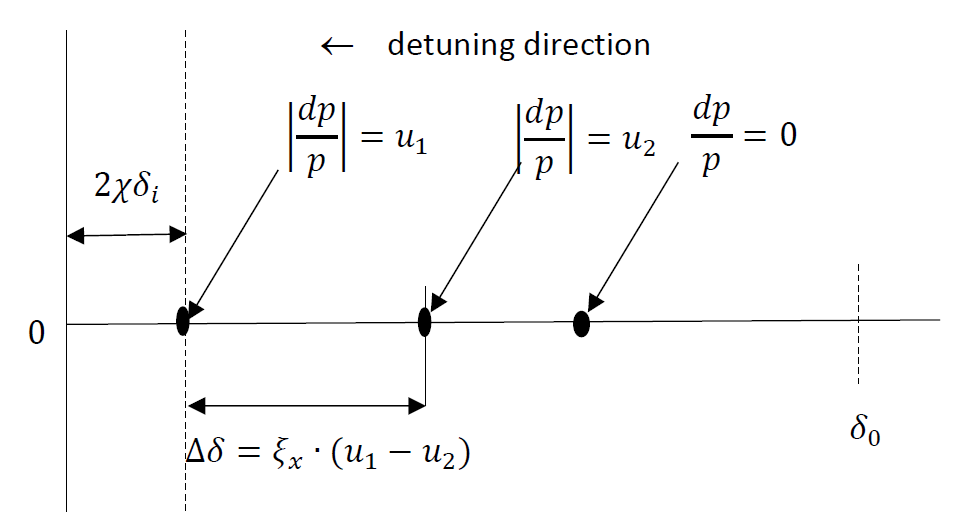}
\caption{
The positions of an equilibrium particle and two particles with nonzero amplitudes of synchrotron oscillations are shown on the resonance detuning scale.}
\label{f7}
\end{figure}

In Fig.6, the spill diagram at right describes the case of the monochromatic extraction. In contrast to the case of no dozed acceleration (the left curve), the extraction with acceleration ends not at the moment when the resonant detuning achieves 0 ($\tilde\Theta=0.5$), but at $\tilde\Theta=1$ (the finish of the detuning change process).
  \begin{figure}
\centering
\includegraphics*[width=90mm]{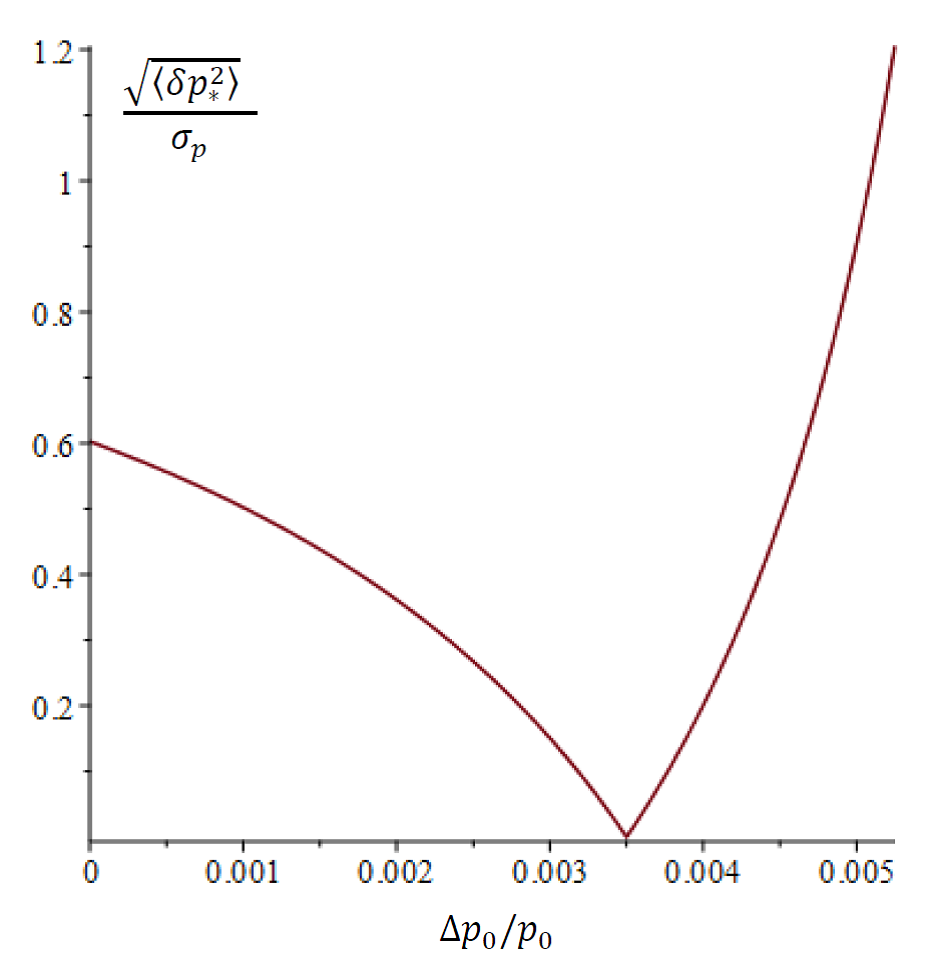}
\caption{The ratio of the momentum spread in extracted beam to that in circulating one depending on the momentum increment due to the additional acceleration at $\xi_x=10$ and $\delta_0=0.035$.}
\label{f8}
\end{figure}

A numerical example of the dependence of the ratio of the momentum spreads of the extracted and circulating beams on the magnitude of the additional acceleration is shown in Fig.8. At $\Delta p_0\rightarrow 2p_0\delta_0/\xi_x$ (i.g. $q \rightarrow 0$) a time to reach the exact resonance becomes infinitely large.

\indent
\section{Experiment at VEPP-4M}
At the storage ring collider VEPP-4M  the experiment \cite{JETPH} was performed at 1.85 GeV to study the behavior of the electron beam near the nolinear resonance $3\nu_y=23$, where $\nu_y$ is the vertical betatron tune. It was a crossing of the resonance in various  modes. In one of the modes  the effective nonlinearity related to the octupole lens was tuned to be close to zero. In this case, no additional stability islands in the phase space  emerge. Besides, the vertical chromaticity was rather low: $\xi_y\approx 0.5 $. In such conditions
one can estimate a law of the particle loss change in a time ($t$) using the results of the adiabatic theory  as applied for the case of vertical oscillations and zero energy spread.  In terms of slow extraction, the time-dependent electron current is described by the equation (4.10) in which the index $x$  is replaced by $y$:
$${\cal J}(t)= \frac{eN}{T_{ex}}\frac{9\pi\delta_0^2}{|{\cal P}|^2\Sigma_y}
\left |1-\frac{2t}
{T_{ex}}\right |\exp{\left\{-\frac{9\pi\delta_0^2 (1-\frac{2t}{T_{ex}})^2}{4|{\cal P}|^2\Sigma_y}\right\}}. \eqno{(4.25)}$$
Here,
$${\cal P}=\left<\frac{\imath h}{4}
\tilde f_y^3 e^{-\imath 23 \theta}\right>,$$
the strength  of the resonance; $h=(\partial^2 B_x/\partial y^2)$, the skew sextupole perturbation in units of an average guide field;  $\tilde f_y$, the conjugated vertical  Floquet function;
$\Sigma_y$, the vertical emittance multiplied by a factor of $2\pi$. 
Regading our experiment,  we replace the current of extracted beam with the rate of beam losses due to crossing the resonance in arbitrary units. 
Beam losses were measured using a synchrotron radiation sensor and a scintillation counter moved in the vertical plane. 

To adjust the required quadratic nonlinearity, it was planned to use the existing skew sextupole lens with an estimated maximum resonance strength of about $ 0.06$ mm$^{-1/2}$ at the energy of experiment. However, as it was established in the course of the experiment, in general, the strength of the resonance was determined by the interference of the contributions from that skew sextupole and from the chromaticity correction system in the arcs \cite{JETPH}. As numerical simulation shows, the contribution of chromatic correction is related to  inclination of the axes of the eigenmodes of the transverse oscillations with respect to the ideal directions along the vertical and horizontal due to the finite betatron coupling\footnote{Recent experiments at VEPP-4M on resonant excitation of transverse oscillations with their registration at turn-by-turn pickup  stations revealed the form of eigenmodes in the form of ellipses with different inclinations of the axes, reaching a few tens of degrees depending on the coupling strength.}.
The coupling effect was enhanced using an skew-quadrupole corrector. A significant part of the experiments was performed without the use of the skew-sextupole lens. These are the results discussed below.

\begin{figure}
\centering
\includegraphics*[width=120mm]{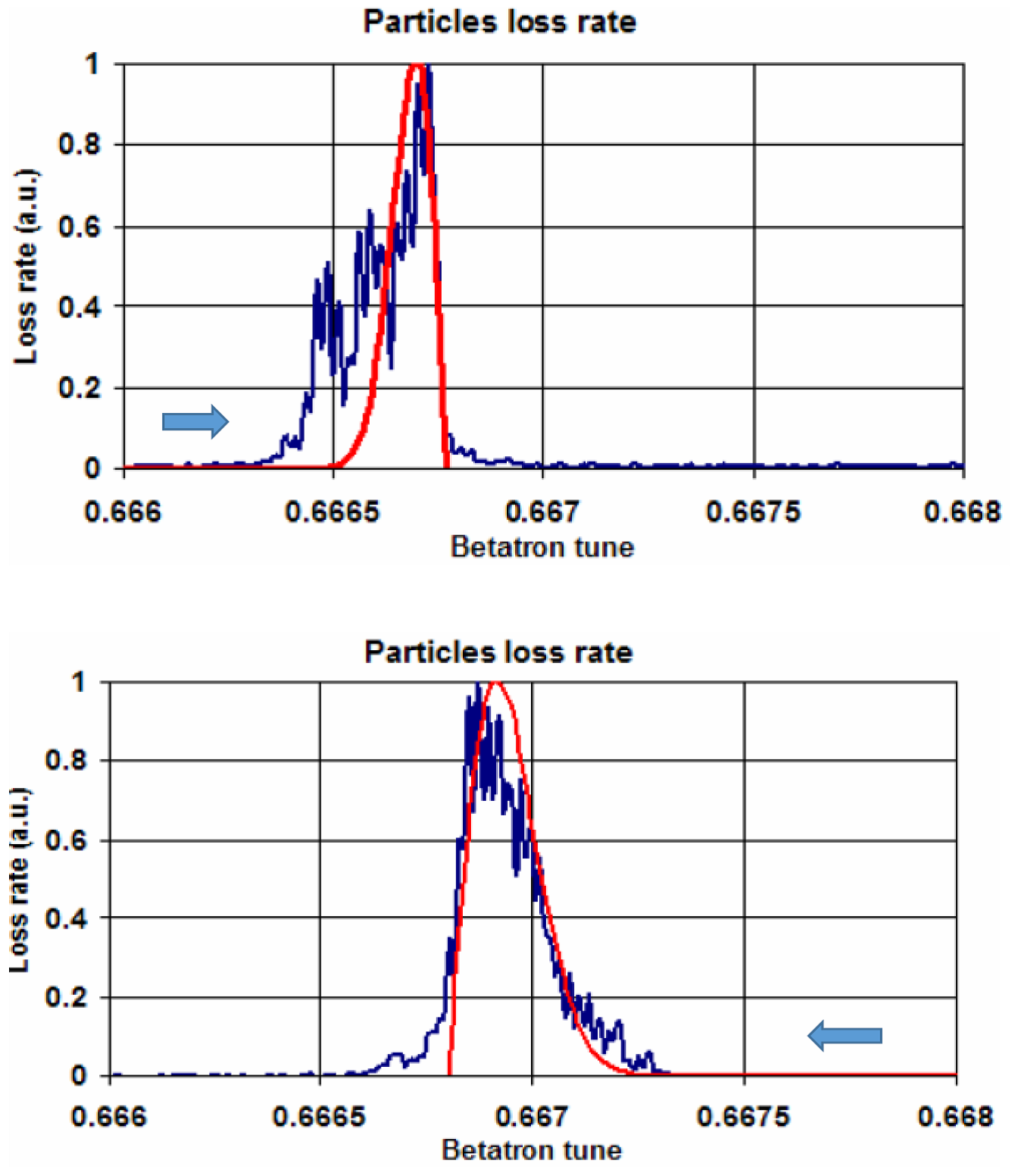}
\caption{ Measured (in black) and calculated (in red) normalized losses of particles in the process of crossing the resonance from bottom to top when changing the frequency for 10 sec in the interval $\nu_y=0.6652 \rightarrow 0.6687$ ((top figure), and from top to bottom for the same time in the interval $\nu_y=0.6685 \rightarrow 0.6653$ (bottom figure).}
\label{f9}
\end{figure}

The resonant harmonic was estimated in the isolated resonance approximation from the distortion of phase trajectories, as well as from the value of the dynamic aperture near the resonance at a certain   detuning \cite{JETPH,Smal}. To do this, the impact of the inflector excited vertical oscillations of the beam, and according to the data of the pickup station, either a change in their amplitude between the maximum and minimum values was observed, or the critical value of the amplitude was fixed at which the beam was lost during a certain number of turns.
It was estimated that the amplitude of the resonant harmonic could lie in the range ($0.02\div0.07$) mm$^ {-1/2}$  or ($0.06 \div 0.22$) cm$^ {-1/2}$ in other units \cite{JETPH}. In the described experiment, an old electronic system with a single-turn pickup resolution of 100 $\mu$m was used\footnote{In 2013–2016, new pickup signal processing modules with a higher accuracy were developed and installed at the VEPP-4M storage ring \cite{Bekh}.}, while the amplitude of the oscillations excited by the impact was fractions of a millimeter. Therefore, the accuracy of perturbation harmonic estimation was quite low.  For this reason, when comparing the experiment with the calculation by formula (4.25), we choose the parameter ${\cal P}$ from the conditions of its closeness to the specified range and the best fit to the measured data . 

 Fig. 9 shows the examples of comparing theory with experiment with a significant coupling and relatively slow tuning of the frequency (by 0.003 per 10 s), in one case, towards its increase, in the other, towards decrease\footnote{A similar figure is shown in \cite{JETPH}, but the directions of detuning change are confused in the caption.}.  The vertical emittance is estimated at the level of 5$\%$  of the calculated horizontal one, which is $2.5\cdot 10^{-6}$ cm $\cdot$ rad at 1850 MeV. When tuning from the bottom to up, the harmonic ${\cal P}= 0.3$  cm$^{-1/2}$  was chosen, when tuning down ${\cal P}= 0.5 $cm$^{-1/2}$ . The jagged structure of the measured  particle loss graphs is apparently associated with the instability (albeit small - on the order of $10^{-4}$) of the power sources of the quadrupole magnets. The given estimates of the harmonic  \cite{JETPH} refer to the definition of the Hamiltonian in the form $H=\delta \cdot I_y+ I_y^{3/2} A_3 \cos{3 w}$,  where $A_3 = 2^{3/2}|{\cal P}|/3\approx |{\cal P}| $.
At $\delta_0=0.003$, $T=10$ seconds, $|{\cal P}|=0.3 $ cm$^{-1/2}$, $\Sigma_y=2\pi\cdot 0.05 \cdot 2.5\cdot 10^{-6}$ cm$\cdot$ rad the adiabatic condition is fulfilled with a large margine:
$$\frac{d\delta}{d t}<< \frac{\omega_0}{9}|{\cal P}|^2 \Sigma_y ~ ~ \rightarrow ~ ~0.003 ~ s^{-1} << 0.041~  s^{-1} . $$

Other loss diagrams obtained in a series of adiabatic resonance crossing, as a rule,  had a similar asymmetric shape -  with a sharper leading edge at the resonance point, as in Fig. 9, in accordance with the prediction of the theory for the case of low chromaticity. 
At the same time, the loss of particles was small - about 10\% (the insertion depth of the scintillation counter was 7.3 mm into a chamber with a diameter of about 8 cm). This contradicts the theory which implies the total loss of particles.  Radiative damping (disregarded by the theory) and residual  cubic nonlinearity for particles with large oscillation amplitudes can be considered as an explanation of this fact

\section{Discussion}
For comparison, let us briefly note other cases in the experiment on crossing a third-order resonance at VEPP-4M, which differ from those considered above.
If the condition of adiabaticity is violated, for example, by reducing the rate of change in detuning by an order of magnitude (to thousandths of 1 s or less) or by reducing the controlled coupling of oscillations, the shape of the loss diagrams becomes close to symmetrical.  In addition, the fraction of lost particles dropped noticeably.
At a high chromaticity value ($\xi_y=9.5$) , in addition to the main bell-shaped distribution (without sharper leading edge), the similar, but lower amplitude distributions of losses on modulation resonances with the frequency of synchrotron oscillations appeared. Since, according to the definition (4.18 ), the adiabaticity condition is deeply violated, this case is also not described by the theory.
In the case of low chromaticity, it can be concluded that the shape of the loss distribution predicted in the theory is in good agreement with the experiment. At the same time, the quantitative verification of the theory is not yet sufficiently precise.
In this regard, it would be interesting to repeat the experiment with the use of a new existing system of turn-to-turn measurements, which will provide a higher accuracy in determination of the harmonic amplitude compared to that achieved earlier.A complete verification of the theory is apparently possible at some ion synchrotron, in which the adiabaticity condition can also be satisfied with allowance for synchrotron oscillations.

\section{Acknowledgements}
The author thanks A.M.~Kondratenko for his interest and some remarks made at the initial stage of the theoretical part of the work.
The author  is grateful to all colleagues, participants of the experiment at VEPP-4M \cite{JETPH}, especially,  E.B.~ Levichev and P.A.~ Piminov - for useful discussions. Special thanks to N.A. Vinokurov and V.A. Vostrikov, who made valuable comments on the manuscript.
 
\end{document}